\documentclass[a4paper,12pt]{article}
\usepackage{latexsym}
\usepackage{amsmath}
\usepackage{amssymb}
\usepackage{amscd}

\newlength{\minitwocolumn}
\makeatletter
\@addtoreset{equation}{section}
\makeatother

\title{\bf 
Finite $XXZ$ critical chain \\
with double boundaries
}

\author{Takeo Kojima}
\date{\it
Department of Mathematics,
College of Science and Technology,\\
Nihon University, Chiyoda-ku, Tokyo
101-0062, Japan\\~\\
{\rm \today}
}

\begin{document}
\maketitle

\begin{abstract}
Finite $XXZ$ chain with double
boundaries is considered at critical regime $-1<\Delta<1$.
We construct
the eigenvectors of finite Hamiltonian by means of
the vertex opertors and the quasi-boundary states.
Using the free field realizations of
the vertex operators and the quasi-boundary states,
integral representations for the correlation functions
are derived.
\end{abstract}

~\\

\section{Introduction}
The $XXZ$ chain is a fundamental model
in understanding of the integrable systems.
Many attentions 
have been 
paid to the $XXZ$ integrable systems
\cite{JM, Bax}.
The purpose of this paper is
to derive correlation functions
for finite $XXZ$ chain with double
boundaries at critical regime
$-1<\Delta<1$,
by means of the free field approach.

In the earlier works 
\cite{JKKKM, JKKMW}
the $XXZ$ chain with a boundary
was considered at massive
regime $\Delta<-1$,
in the framework of the free field approach.
The integral representations
of the correlation functions were
derived.
It was shown that boundary quantum
Knizhnik-Zamolodchikov equations
with the certain shift,
governed the correlation functions.
The $U_q(\widehat{sl_n})$-generalization
of the papers on the $XXZ$ chain \cite{JKKKM, JKKMW}
was given in \cite{FK, KQ}.
In the paper
\cite{K}
results for finite $XXZ$ chain at massive regime
$\Delta<-1$
were extended to critical
regime $-1<\Delta<1$,
using bosonizations of 
vertex operators \cite{JKM}.

Y.Fujii and M.Wadati \cite{FW}
noticed that 
solutions of boundary quantm Knizhnik-
Zamolodchikov equations without shift
became eigenstates of
finite $XXZ$ chain with double boundaries.
They constructed eigenstates
of finite $XXZ$ chain with double boundaries
at massive regime $\Delta<-1$,
by means of the free field approach.
In the present paper
we shall 
study finite $XXZ$ chain with double boundaries
at
critical regime $-1<\Delta <1$.
We shall derive the correlation functions as
integrals of meromorphic functions involving
Multi-Gamma functions.

Now a few words about organiozation
of this paper.
In section 2 we formulate the problem.
In section 3 we construct
the realizations of 
eigenstates.
In section 4 we derive 
integral representations for 
the correlation functions.
In Appendix A we summarized the bosonizations
of the vertex operators \cite{JKM}.
In Appendix B we summarized the Multi-Gamma
functions.

\section{Boundary quantum KZ-equation}
In 1984 I.V.Cherdnik \cite{C} 
proposed the following systems of difference equations,
now called boundary quantum Knizhnik-Zamolodchikov equations.
\begin{eqnarray}
&&F(\beta_1,\cdots,\beta_j+i\lambda,\cdots,\beta_N)
\nonumber\\
&=&T_j(\beta_1,\cdots,\beta_N|\lambda)
F(\beta_1,\cdots,\beta_j,\cdots,\beta_N),
~(j=1,\cdots,N),
\label{BQKZ} 
\end{eqnarray}
where the shift operator
$T_j(\beta_1 \cdots \beta_N|\lambda)$
is given by
\begin{eqnarray}
T_j(\beta_1,\cdots,\beta_N|\lambda)&=&
R_{j,j-1}(\beta_j-\beta_{j-1}+i\lambda)\cdots
R_{j,1}(\beta_j-\beta_1+i\lambda)\bar{K}_j(\beta_j)\nonumber\\
&\times&
R_{1,j}(\beta_1+\beta_j)\cdots R_{j-1,j}(\beta_{j-1}+\beta_j)
R_{j+1,j}(\beta_{j+1}+\beta_j)\cdots R_{N,j}(\beta_{N}+\beta_j)
\nonumber\\
&\times&K_j(\beta_j)
R_{j,N}(\beta_j-\beta_N)\cdots
R_{j,j+1}(\beta_j-\beta_{j+1}).
\end{eqnarray}
The $R$-matrix $R(\beta)$ and the boundary $K$-matrix
$K(\beta), \bar{K}(\beta)$ are specified later.
The solutions of the boundary quantum KZ equations
represent various physical quantities.
For the case of the shift parameter $\lambda=2\pi$,
the certain solutions of the quantum KZ equations 
represents $N$-point correlation functions
for the massless $XXZ$ chain with a boundary \cite{K},
which is described by the following Hamiltonian :
\begin{eqnarray}
{\cal H}=-\frac{1}{2}
\sum_{n=1}^{\infty}
\left(\sigma_{n+1}^x\sigma_n^x+
\sigma_{n+1}^y\sigma_n^y+\Delta \sigma_{n+1}^z\sigma_n^z
\right)+h \sigma_1^z,
\end{eqnarray}
where we set a parameter $-1<\Delta<1$ and
a parameter $h$ represents the boundary external field.
The $\sigma_n^x,\sigma_n^y$ and $\sigma_n^z$
stand for the Pauli matrices acting on
the $n$-th site of the half {\bf Infinite} spin chain :
$\cdots \otimes {\mathbb C}^2 
\otimes {\mathbb C}^2 \otimes {\mathbb C}^2$. 
The author \cite{K} derived the integrale representations
of the correlation functions for the above model.
\\
In the present paper we shall consider the case of 
the shift parameter $\lambda=0$.
In this case the solution of the boundary
quantum KZ equation (\ref{BQKZ}) represents
the eigenvector of finite $XXZ$ chain with double boundaries
at critical regime $-1<\Delta<1$.
The Hamiltonian ${\cal H}_F$ of our considering model
is given by
\begin{eqnarray}
{\cal H}_F=-\frac{1}{2}
\sum_{n=1}^{N-1}
\left(\sigma_{n+1}^x\sigma_n^x+
\sigma_{n+1}^y\sigma_n^y+\Delta \sigma_{n+1}^z\sigma_n^z
\right)+h_1 \sigma_1^z +h_N \sigma_N^z,
\label{Hamiltonian}
\end{eqnarray}
where we set a parameter $-1<\Delta<1$.
Parameters $h_1, h_N$ represent the boundary external fields.
The $\sigma_n^x,\sigma_n^y$ and $\sigma_n^z$
stand for the Pauli matrices acting on
the $n$-th site of the {\bf Finite} spin chain :
${\left({\mathbb C}^2\right)}^{\otimes N}$.

~\\
Let us set the $R$-matrix as
\begin{eqnarray}
R(\beta)=r(\beta)
\left(\begin{array}{cccc}
1&&&\\
&b(\beta)
&
c(\beta)
&\\
&c(\beta)
&
b(\beta)
&\\
&&&1
\end{array}
\right), \label{def:R}
\end{eqnarray}
where we set the components as
\begin{eqnarray}
b(\beta)=-\frac{\displaystyle {\rm sh}\left(\frac{\beta}{\xi+1}\right)}
{\displaystyle {\rm sh}\left(\frac{\beta+\pi i}{\xi+1}\right)
}~~~,
c(\beta)=\frac{\displaystyle {\rm sh}\left(\frac{\pi i}{\xi+1}\right)}
{\displaystyle {\rm sh}\left(\frac{\beta+\pi i}{\xi+1}\right)
}.
\end{eqnarray}
Here we set
\begin{eqnarray}
r(\beta)=-
\frac{S_2(i\beta|2\pi,\pi(\xi+1))
S_2(-i\beta+\pi|2\pi,\pi(\xi+1))}
{S_2(-i\beta|2\pi,\pi(\xi+1))
S_2(i\beta+\pi|2\pi,\pi(\xi+1))
},
\end{eqnarray}
where $S_2(\beta|\omega_1 \omega_2)$ is the double sine 
function defined in Appendix B.
\\
Let $\{v_+,v_-\}$ denote the natural basis
of $V={\mathbb{C}}^2$.
When viewed as an operator on
$V \otimes V$, the matrix elements of $R(\beta)$
are defined by
\begin{eqnarray}
R(\beta)v_{k_1}\otimes v_{k_2}=
\sum_{j_1,j_2=\pm}
v_{j_1}\otimes v_{j_2}R(\beta)_{j_1 j_2}^{k_1 k_2}.
\end{eqnarray}
The $R$-matrix satisfies the Yang-Baxter equation :
\begin{eqnarray}
R_{12}(\beta_1-\beta_2)R_{13}(\beta_1-\beta_3)
R_{23}(\beta_2-\beta_3)=
R_{23}(\beta_2-\beta_3)R_{13}(\beta_1-\beta_3)
R_{12}(\beta_1-\beta_2).
\end{eqnarray}
The normalization factor $r_0(\beta)$ is so chosen that
the unitarity and crossing relations are
\begin{eqnarray}
R_{12}(\beta)R_{21}(-\beta)&=&id,\\
R(-\beta)_{j_1 j_2}^{k_1 k_2}&=&
R(\beta-\pi i)_{-k_2 j_1}^{-j_2 k_1}.
\end{eqnarray}
Let us set the boundary $K$-matrix $K(\beta)$ by
\begin{eqnarray}
K(\beta)=k(\beta)\left(\begin{array}{cc}
1&0\\
0&
\frac{\displaystyle {\rm sh}\left(\frac{\nu+\beta}{\xi+1}\right)}
{\displaystyle
{\rm sh}\left(\frac{\nu-\beta}{\xi+1}\right)
}
\end{array}\right),\label{def:K}
\end{eqnarray}
where the normalization factor is given by
\begin{eqnarray}
k(\beta)=k_0(\beta)k_1(\beta),
\end{eqnarray}
where
\begin{eqnarray}
k_0(\beta)=\frac{S_2(-2i\beta+4\pi|4\pi,\pi(\xi+1))
S_2(2i\beta+3\pi|4\pi,\pi(\xi+1))
}{
S_2(2i\beta+4\pi|4\pi,\pi(\xi+1))
S_2(-2i\beta+3\pi|4\pi,\pi(\xi+1))
}
,\\
k_1(\beta)=\frac{
S_2(-i\beta+i\nu+\pi|2\pi,\pi(\xi+1))
S_2(i\beta+i\nu+2\pi|2\pi,\pi(\xi+1))
}{
S_2(i\beta+i\nu+\pi|2\pi,\pi(\xi+1))
S_2(-i\beta+i\nu+2\pi|2\pi,\pi(\xi+1))
}
.
\end{eqnarray}
The matrix elements $K(\beta)_{j}^k$ are
defined by
\begin{eqnarray}
K(\beta)v_k=\sum_{j=\pm}v_j K(\beta)_j^k.
\end{eqnarray}
The $R$-matrix and the $K$-matrix satisfy the Boundary
Yang-Baxter equation.
\begin{eqnarray}
K_2(\beta_2)R_{21}(\beta_1+\beta_2)K_1(\beta_1)
R_{12}(\beta_1-\beta_2)=
R_{21}(\beta_1-\beta_2)K_1(\beta_1)R_{12}(\beta_1+\beta_2)
K_2(\beta_2).
\end{eqnarray}
The normalization factor $k(\beta)$ is so chosen 
that
the boundary unitarity and the boundary crossing relations are
\begin{eqnarray}
K(\beta)K(-\beta)&=&id,\\
K\left(\beta+\frac{\pi i}{2}\right)_j^j&=&
\sum_{k=\pm}R(2\beta)_{k,-k}^{-j,j}
K\left(-\beta+\frac{\pi i}{2}\right)_k^k.
\end{eqnarray}
Let us set the boundary $K$-matrix $\bar{K}(\beta)$ by
\begin{eqnarray}
\bar{K}(\beta)=K(\beta)|_{\mu \leftrightarrow \nu}.
\end{eqnarray}
{\it Note. ~For the another shift parameter $\lambda=2\pi$ case,
we take another choice of the $K$-matrix $\bar{K}(\beta)$.
See the reference \cite{K}.}

~\\
The derivatives of $R$-matrix and $K$-matrix are given by
\begin{eqnarray}
&&\left.\frac{\partial}{\partial \beta}R_{j,j+1}(\beta)P_{j,j+1}
\right|_{\beta=0}\nonumber\\
&=&
\frac{-1}{2(\xi+1)}\times \frac{1}{\displaystyle
{\rm sh}\left(
\frac{\pi i}{\xi+1}\right)}
\left(\sigma_j^x \sigma_{j+1}^x +
\sigma_j^y \sigma_{j+1}^y +\Delta \sigma_{j}^z \sigma_{j+1}^z
\right)+const.
\end{eqnarray}
\begin{eqnarray}
\left.\frac{\partial}{\partial \beta}K_{j}(\beta)
\right|_{\beta=0}=\frac{-1}{\xi+1}
\times\frac{\displaystyle {\rm ch}
\left(\frac{\nu}{\xi+1}\right)}{
\displaystyle
{\rm sh}
\left(\frac{\nu}{\xi+1}\right)
}\sigma_j^z+const.
\end{eqnarray}
Here the anisotropic parameter $\Delta=-{\rm cos}
\left(\frac{\pi}{\xi+1}\right)$.\\
Therefore the shift operator $T_j(\beta_1 \cdots \beta_N|0)$
is related to the Hamiltonian ${\cal H}_F$ (\ref{Hamiltonian})
as following.
\begin{eqnarray}
{\rm sh}\left(\frac{\pi i}{\xi+1}\right)\times
\frac{\xi+1}{2}\times
\left(\frac{\partial}{\partial \beta_j}T_{j}\right)
(0,\cdots,0|0)={\cal H}_F + const.
\end{eqnarray}
Here the boundary magnetic fields $h_1, h_N$ are
related with parameters $\mu, \nu$,
\begin{eqnarray}
h_1=-\frac{1}{2}\times{\rm sh}\left(\frac{\pi i}{\xi+1}\right)
\frac{\displaystyle{\rm ch}\left(\frac{\nu}{\xi+1}\right)}{
\displaystyle{\rm sh}\left(\frac{\nu}{\xi+1}\right)
},~~
h_N=-\frac{1}{2}\times{\rm sh}\left(\frac{\pi i}{\xi+1}\right)
\frac{\displaystyle{\rm ch}\left(\frac{\mu}{\xi+1}\right)}{
\displaystyle
{\rm sh}\left(\frac{\mu}{\xi+1}\right)
}.
\end{eqnarray}
We have 
\begin{eqnarray}
T_j(0,\cdots,0|0)=id.
\end{eqnarray}
Let us set the eigenvector $|\beta_1 \cdots \beta_N\rangle$
by
\begin{eqnarray}
T_j(\beta_1,\cdots,\beta_N|0)|\beta_1,\cdots,\beta_N\rangle
=|\beta_1,\cdots,\beta_N\rangle.
\end{eqnarray}
Let us set the dual eigenvector by
\begin{eqnarray}
\langle \beta_1,\cdots, \beta_N|
T_j(\beta_1,\cdots,\beta_N|0)
=\langle \beta_1,\cdots,\beta_N|.
\end{eqnarray}
The above eigenvectors satisfy
the followings.
\begin{eqnarray}
{\cal H}_F |0,\cdots,0\rangle=
Const.|0,\cdots,0\rangle,~~
\langle 0,\cdots,0|{\cal H}_F=
Const.\langle 0,\cdots,0|.
\end{eqnarray}
In the next section we shall construct
the eigenvector
$|\beta_1 \cdots \beta_N\rangle$ explicitly.

\section{Eigenvectors}
In this section we solve following
eigenvector problem.
\begin{eqnarray}
{\cal T}_j(\beta_1,\cdots,\beta_N|0)
|\beta_1,\cdots, \beta_N\rangle=
|\beta_1,\cdots, \beta_N\rangle.\label{eigen}
\end{eqnarray}
The eigenvector is realized by using
the vertex operators $\Phi_j(\beta)$.
\begin{eqnarray}
|\beta_1,\cdots,\beta_N\rangle=
\frac{1}{\langle G|G\rangle}
\sum_{\epsilon_1 \cdots \epsilon_N=\pm}
\langle G|\Phi_{\epsilon_1}(\beta_1) 
\cdots \Phi_{\epsilon_N}(\beta_N)|G\rangle
(v_{\epsilon_1}\otimes \cdots \otimes v_{\epsilon_N}).
\end{eqnarray}
Here the vector $|G\rangle$ and the dual vector $\langle G|$
are characterized by the following relations.
\begin{eqnarray}
\langle G|\Phi_j(\beta)=
\bar{K}(\beta)_j^j\langle G|\Phi_j(-\beta),~(j=\pm),
\label{characterI}
\\
\Phi_j(-\beta)|G\rangle=
K(\beta)_j^j\Phi_j(\beta)|G\rangle,~(j=\pm),
\label{characterII}
\end{eqnarray}
We call the auxiliary states $\langle G |$ and $|G\rangle$
``quasi-boundary state''.
Using the following 
commutation relation of the vertex operator and
the characterizing relations (\ref{characterI}, \ref{characterII})
of the quasi-boundary state,
we have the equation (\ref{eigen}).
\begin{eqnarray}
\Phi_{j_1}(\beta_1)\Phi_{j_2}(\beta_2)=
\sum_{k_1,k_2=\pm}R(\beta_1-\beta_2)
_{j_1,j_2}^{k_1,k_2}
\Phi_{k_2}(\beta_2)\Phi_{k_1}(\beta_1).
\end{eqnarray}

~\\
We give the free field realization of
the quasi-boundary state.\\
Let us introduce the free bosons $b(t), (t \in {\mathbb R})$ by
\begin{eqnarray}
[b(t),b(t')]=\frac{
\displaystyle
{\rm sh}\left(\frac{\pi t}{2}\right)
{\rm sh}(\pi t){\rm sh}\left(\frac{\pi t \xi}{2}\right)
}{t
\displaystyle 
{\rm sh}\left(\frac{\pi t (\xi+1)}{2}\right)
}\delta(t+t').
\end{eqnarray}
Let us set the Fock space ${\cal H}$ generated by
the vacuum vector $\langle vac |$ which satisfies
\begin{eqnarray}
\langle vac |b(-t)=0,~{\rm if}~t>0.
\end{eqnarray}
The quasi-boundary state $\langle G |$ is realized
as followings.
\begin{eqnarray}
\langle G|=\langle vac |e^G,
\end{eqnarray}
Here we have set
\begin{eqnarray}
G=\frac{1}{2}\int_0^\infty
\frac{G_2(t|\mu)}{[b(t),b(-t)]}b(t)^2 dt+
\int_0^\infty \frac{G_1(t|\mu)}{[b(t),b(-t)]}b(t) dt,
\end{eqnarray}
where
\begin{eqnarray}
&&G_2(t|\mu)=-1,\\
&&G_1(t|\mu)=
\frac{1}{t}
\frac{\displaystyle
{\rm sh}\left(\frac{\pi}{2}t\right)
{\rm sh}\left((-i\mu+\frac{\pi}{2}\xi)t\right)
}{\displaystyle
{\rm sh}\left(\frac{\pi}{2}(\xi+1)t\right)
}
+\frac{1}{t}
\frac{\displaystyle
{\rm sh}\left(\frac{\pi}{4}t\right)
{\rm sh}\left(\frac{\pi}{2}t\right)
{\rm ch}\left(\frac{\pi}{4}\xi t\right)
}{\displaystyle
{\rm sh}\left(\frac{\pi}{4}(\xi+1) t\right)
}.
\end{eqnarray}
Let us prove the relation (\ref{characterI}).
In what follows we use the abberiviations.
\begin{eqnarray}
U_+(\beta)=
\exp\left(-\int_0^\infty
\frac{b(t)}{{\rm sh}\pi t}e^{i\beta t}dt\right),
U_-(\beta)=
\exp\left(\int_0^\infty
\frac{b(-t)}{{\rm sh}\pi t}e^{-i\beta t}dt\right),\\
\bar{U}_+(\alpha)=
\exp\left(\int_0^\infty
\frac{b(t)}{{\rm sh}\frac{\pi}{2} t}e^{i\alpha t}dt\right),
\bar{U}_-(\alpha)=
\exp\left(-\int_0^\infty
\frac{b(-t)}{{\rm sh}\frac{\pi}{2} t}e^{-i\alpha t}dt\right).
\end{eqnarray}
In what follows we omit non-essential constant factors.\\
At first we explain the formulas of the form
\begin{eqnarray}
X(\beta_1)Y(\beta_2)=C_{XY}(\beta_1-\beta_2)
:X(\beta_1)X(\beta_2):,
\end{eqnarray}
where $X,Y=U_j$, and $C_{XY}(\beta)$ is a meromorphic
function on ${\mathbb{C}}$.
These formulae follow from the commutation relation
of the free bosons.
When we compute the contraction of the basic
operators,
we often encounter an integral
\begin{eqnarray}
\int_0^\infty
F(t)dt,
\end{eqnarray}
which is divergent at $t=0$.
Here we adopt the following prescription
for regularization :
it should be understood as the countour integral,
\begin{eqnarray}
\int_C F(t)\frac{{\rm log}(-t)}{2\pi i}dt, 
\end{eqnarray}
where the countour $C$ is given by
\\
~\\
~\\
\unitlength 0.1in
\begin{picture}(34.10,11.35)(17.90,-19.35)
%
\special{pn 8}%
\special{pa 5200 800}%
\special{pa 2190 800}%
\special{fp}%
\special{sh 1}%
\special{pa 2190 800}%
\special{pa 2257 820}%
\special{pa 2243 800}%
\special{pa 2257 780}%
\special{pa 2190 800}%
\special{fp}%
\special{pa 2190 1600}%
\special{pa 5190 1600}%
\special{fp}%
\special{sh 1}%
\special{pa 5190 1600}%
\special{pa 5123 1580}%
\special{pa 5137 1600}%
\special{pa 5123 1620}%
\special{pa 5190 1600}%
\special{fp}%
%
\special{pn 8}%
\special{pa 5190 1200}%
\special{pa 2590 1210}%
\special{fp}%
\put(25.9000,-12.1000){\makebox(0,0)[lb]{$0$}}%
%
\special{pn 8}%
\special{ar 2190 1210 400 400  1.5707963 4.7123890}%
\put(33.9000,-20.2000){\makebox(0,0){{\bf Contour} $C$}}%
\end{picture}%

~\\
The actions of the basic oprtaors on quasi-boundary state
$\langle G|$ are evaluated as followings.
\begin{eqnarray}
\langle G |U_-(\beta)=Const.m(\beta)
\langle G |U_+(-\beta),\\
\langle G |\bar{U}_-(\alpha)=Const.J(\alpha)
\langle G |\bar{U}_+(-\alpha).
\end{eqnarray}
Here we have set
\begin{eqnarray}
m(\beta)&=&\frac{\Gamma_2(2i\beta+4\pi|2\omega_1 \omega_2)
\Gamma_2(2i\beta+\pi(\xi+1)|2\omega_1 \omega_2)
}{
\Gamma_2(2i\beta+3\pi|2\omega_1 \omega_2)
\Gamma_2(2i\beta+\pi(\xi+2)|2\omega_1 \omega_2)
}
\nonumber
\\
&\times&\frac{
\Gamma_2(i\beta+i\mu+\pi|\omega_1 \omega_2)
\Gamma_2(i\beta-i\mu+\pi(\xi+2)|\omega_1 \omega_2)
}{
\Gamma_2(i\beta+i\mu+2\pi|\omega_1 \omega_2)
\Gamma_2(i\beta-i\mu+\pi(\xi+1)|\omega_1 \omega_2)
},\\
J(\alpha)&=&\alpha \times 
\frac{\Gamma\left(\frac{-i\mu+i\alpha}{\pi(\xi+1)}+1
-\frac{1}{2(\xi+1)}\right)}{
\Gamma\left(\frac{i\mu+i\alpha}{\pi(\xi+1)}
+\frac{1}{2(\xi+1)}\right)
}.
\end{eqnarray}
We have
\begin{eqnarray}
\langle G |\Phi_+(\beta)=
m(\beta)\langle G |U_+(\beta)U_+(-\beta).
\end{eqnarray}
Because the function $m(\beta)$ satisfies
\begin{eqnarray}
\bar{K}(\beta)_+^+=\frac{m(\beta)}{m(-\beta)},
\end{eqnarray}
we have proved the $''+''$-part of the characterizing
relation (\ref{characterI}).\\
We will prove the $''-''$-part of the equation 
(\ref{characterI}).
Using the actions formulae of the basic operators on
the quasi-boundary state, we have the following.
\begin{eqnarray}
\langle G|\Phi_-(\beta)
&=&Const.
m(\beta)\int_{-\infty}^\infty
d\alpha \times \alpha \times 
\prod_{\epsilon_1,\epsilon_2=\pm}
\Gamma\left(\frac{i(\epsilon_1 \alpha +\epsilon_2 \beta)}
{\pi(\xi+1)}+\frac{1}{2(\xi+1)}\right)\nonumber\\
&\times&{\rm sh}\left(\frac{\alpha+\beta}{\xi+1}+
\frac{\pi i}{2(\xi+1)}\right)
\frac{\Gamma\left(\frac{-i\mu+i\alpha}{\pi(\xi+1)}
+1-\frac{1}{2(\xi+1)}\right)}{
\Gamma\left(\frac{i\mu+i\alpha}{\pi(\xi+1)}
+\frac{1}{2(\xi+1)}\right)
}\nonumber\\
&\times&\langle G |U_+(\beta)U_+(-\beta)\bar{U}_+(\alpha)
\bar{U}_+(-\alpha).
\end{eqnarray}
Note that the operator part of the above equation :
$\langle G |U_-(\beta)U_-(-\beta)
\bar{U}_-(\alpha)\bar{U}_-(-\alpha)$
is invariant under the changes of variables :
$\alpha \leftrightarrow -\alpha,
\beta \leftrightarrow -\beta$.
\\
We have
\begin{eqnarray}
&&m(\beta)^{-1}
{\rm sh}\left(\frac{\mu-\beta}{\xi+1}\right)
\langle G |\Phi_-(\beta)
-m(-\beta)^{-1}
{\rm sh}\left(\frac{\mu+\beta}{\xi+1}\right)
\langle G |\Phi_-(-\beta)\nonumber\\
&=&
Const.\times
{\rm sh}\left(\frac{2\beta}{\xi+1}\right)
\int_{-\infty}^\infty
d\alpha \prod_{\epsilon_1,\epsilon_2=\pm}
\Gamma\left(\frac{i(\epsilon_1 \alpha +\epsilon_2 \beta)}{
\pi(\xi+1)}+\frac{1}{2(\xi+1)}\right)\nonumber\\
&\times&
\prod_{\epsilon=\pm}\Gamma
\left(\frac{i(-\mu+\epsilon \alpha)}{\pi(\xi+1)}
+1-\frac{1}{2(\xi+1)}\right)\times \alpha
\prod_{\epsilon=\pm}{\rm sh}\left(
\frac{\mu+\epsilon \alpha}{\xi+1}-\frac{\pi i}{2(\xi+1)}
\right)\nonumber\\
&\times& 
\langle G|
U_-(\beta)U_-(-\beta)\bar{U}_-(\alpha)\bar{U}_-(-\alpha).
\end{eqnarray}
The integrand of (RHS) is anti-symmetric to a change
of integral variable $\alpha \leftrightarrow -\alpha$.
It means the left-hand side becomes zero after taking integral.
Therefore we get
\begin{eqnarray}
m(-\beta){\rm sh}\left(\frac{\mu-\beta}{\xi+1}\right)
\langle G|\Phi_-(\beta)=
m(\beta){\rm sh}\left(\frac{\mu+\beta}{\xi+1}\right)
\langle G|\Phi_-(-\beta).
\end{eqnarray}
We have proved the $''-''$-part
of the characterizing relation (\ref{characterI}).\\
The quasi-boundary state $|G\rangle$ is given by
the following.
\begin{eqnarray}
|G\rangle=e^{G*}|vac\rangle,
\end{eqnarray}
where
\begin{eqnarray}
G^*=\frac{1}{2}\int_0^\infty
\frac{G_2^*(t|\mu)}{[b(t),b(-t)]}b(-t)^2 dt
+\int_0^\infty
\frac{G_1^*(t|\mu)}{[b(t),b(-t)]}b(-t) dt,
\end{eqnarray}
where
\begin{eqnarray}
G_2^*(t|\mu)=G_2(t|\mu),~
G_1^*(t|\mu)=-G_1(t|\mu).
\end{eqnarray}
As the same manner as the above we can prove the characterizing
equations (\ref{characterII}).

~\\
{\it Note.~
When we consider the massless XXZ chain with a boundary \cite{K}.
We introduce the boundary state $|B\rangle$ and the dual boundary
state $\langle B|$.
Left quasi-boundary state $\langle G|$ differes from the
dual boundary state $\langle B|$.
Right quasi-boundary state $|G\rangle$ coincides with the
boundary state $|B\rangle$.
Physically
the boundary state $|B\rangle$ of the paper \cite{K}
coresponds to the vacuum expectation value
$\langle G|\Phi(\beta_1)
\cdots \Phi(\beta_N)|G\rangle$ of
the present paper.
Both quantities $|B\rangle$ and
$\langle G|\Phi(\beta_1)
\cdots \Phi(\beta_N)|G\rangle
$ represent an eigenvector of the Hamiltonian for
each model.}

Let us construct the dual stationary state,
\begin{eqnarray}
\langle \beta_1,\cdots \beta_N|T_j(\beta_1,\cdots,\beta_N|0)
=\langle \beta_1,\cdots \beta_N|.
\end{eqnarray}
The dual stationary state is realized by using the
dual vertex operators $\Phi_j^*(\beta)$.
\begin{eqnarray}
\langle \beta_1,\cdots \beta_N|=
\frac{1}{\langle F|F \rangle}
\sum_{\epsilon_1,\cdots,\epsilon_N=\pm}
\langle F|\Phi_{\epsilon_1}^*(\beta_1)\cdots
\Phi_{\epsilon_N}^*(\beta_N)|F\rangle
(v_{\epsilon_1}^* \otimes \cdots \otimes 
v_{\epsilon_N}^*).
\end{eqnarray}
The dual vertex operators
are related to the vertex operators,
\begin{eqnarray}
\Phi_j^*(\beta)=\Phi_{-j}(\beta+\pi i),~(j=\pm).
\end{eqnarray}
The quasi-boundary state $\langle F|$ and $|F\rangle$
are characterized by
\begin{eqnarray}
\langle F|\Phi_j^*(\beta)={K}(\beta)_j^j
\langle F|\Phi_j^*(-\beta),~(j=\pm),\\
\Phi_j^*(-\beta)|F\rangle
=\bar{K}(\beta)_j^j\Phi_j^*(\beta)|F\rangle,~(j=\pm).
\end{eqnarray}
The quasi-boundary states $\langle F|$ and $|F\rangle$
are realized as followings.
\begin{eqnarray}
\langle F|=\langle vac |e^{F},~~|F\rangle
=e^{F*}|vac \rangle.
\end{eqnarray}
Here we have set
\begin{eqnarray}
F&=&\frac{1}{2}\int_0^\infty
\frac{F_2(t|\mu)}{[b(t),b(-t)]}b(t)^2dt
+\int_0^\infty \frac{F_1(t|\mu)}{[b(t),b(-t)]}b(t)dt,\\
F^*&=&\frac{1}{2}\int_0^\infty
\frac{F_2^*(t|\nu)}{[b(t),b(-t)]}b(-t)^2dt
+\int_0^\infty \frac{F_1^*(t|\nu)}{[b(t),b(-t)]}b(-t)dt,
\end{eqnarray}
where 
\begin{eqnarray}
F_2(t|\mu)&=&-e^{-2\pi t},\\
F_1(t|\mu)&=&
\frac{e^{-\pi t}}{t}
\frac{\displaystyle
{\rm sh}\left(\frac{\pi t}{2}\right)
{\rm sh}\left((i\mu-\frac{\pi \xi}{2}-\pi)t\right)
}{\displaystyle
{\rm sh}\left(\frac{\pi}{2}(\xi+1)t\right)}+\frac{e^{-\pi t}}{t}
\frac{\displaystyle
{\rm sh}\left(\frac{\pi t}{2}\right)
{\rm sh}\left(\frac{\pi t}{4}\right){\rm ch}\left(
\frac{\pi}{4}\xi t\right)
}{\displaystyle
{\rm sh}\left(\frac{\pi}{4}(\xi+1)t\right)
}.\nonumber\\
\end{eqnarray}
\begin{eqnarray}
F_2^*(t|\nu)&=&-e^{2\pi t},\\
F_1^*(t|\nu)&=&-e^{2\pi t}\times
F_1(t|\nu).
\end{eqnarray}
The characterizing relations are proved as the same 
manner. Here we omit details.
Formally other eigenvectors are constructed by inserting
the type-II vertex operators.
\begin{eqnarray}
\sum_{\epsilon_1 \cdots \epsilon_N=\pm}
\langle G|\Phi_{\epsilon_1}(\beta_1)
\cdots \Phi_{\epsilon_N}(\beta_N)
\Psi_{j_1}^*(\xi_1)\cdots
\Psi_{j_M}^*(\xi_M)
|G\rangle
(v_{\epsilon_1}\otimes
\cdots \otimes v_{\epsilon_N}).
\end{eqnarray}

\section{Correlation Functions}
In this section we calculate the vacuum expectation values of the
type-I vertex operators, and obtain
them as integrals of meromorphic functions
involving Multi-Gamma functions.
We compute the following $2N$-point function,
\begin{eqnarray}
&&P_{\epsilon_1^*,\cdots,\epsilon_N^*;
\epsilon_N,\cdots,\epsilon_1;\eta}(\{ \beta_j^* \};
\{\beta_j \})\nonumber\\
&=&
\frac{
\langle F_\eta| \Phi_{\epsilon_1^*}(\beta_1^*)
\cdots \Phi_{\epsilon_N^*}(\beta_N^*) |F_\eta\rangle}
{\langle F_\eta | F_\eta\rangle}\times
\frac{\langle G_\eta|
\Phi_{\epsilon_N}(\beta_N)\cdots \Phi_{\epsilon_1}(\beta_1)
|G_\eta \rangle}
{\langle G_\eta | G_\eta
\rangle}.
\end{eqnarray}
Here we set the state $\langle G_\eta|,
|G_\eta \rangle$ and $\langle F_\eta |, |F_\eta\rangle$ by
\begin{eqnarray}
\langle G_\eta|=\langle vac |e^{G_\eta},~~
|G_\eta \rangle=e^{G_\eta^*}|vac\rangle,\\
\langle F_\eta|=\langle vac |e^{F_\eta},~~
|F_\eta \rangle=e^{F_\eta^*}|vac\rangle.
\end{eqnarray}
where
\begin{eqnarray}
G_\eta&=&
\frac{1}{2}
\int_0^\infty \frac{e^{-\eta t}G_2(t|\mu))}{[b(t),b(-t)]}b(t)^2 dt
+\int_0^\infty
\frac{G_1(t|\mu)}{[b(t),b(-t)]}b(t)dt,\\
G_\eta^*&=&
\frac{1}{2}
\int_0^\infty \frac{e^{-\eta t}G_2^*
(t|\mu))}{[b(t),b(-t)]}b(-t)^2 dt
+\int_0^\infty
\frac{G_1^*(t|\mu)}{[b(t),b(-t)]}b(-t)dt
\\
F_\eta&=&
\frac{1}{2}
\int_0^\infty \frac{e^{-\eta t}F_2(t|\nu
)}{[b(t),b(-t)]}b(t)^2 dt
+\int_0^\infty
\frac{F_1(t|\nu)}{[b(t),b(-t)]}b(t)dt,\\
F_\eta^*&=&
\frac{1}{2}
\int_0^\infty \frac{e^{-\eta t}F_2^*(t|\nu
)}{[b(t),b(-t)]}b(-t)^2 dt
+\int_0^\infty
\frac{F_1^*(t|\nu)}{[b(t),b(-t)]}b(-t)dt.
\end{eqnarray}
We have
\begin{eqnarray}
\lim_{\eta \to 0}\langle G_\eta|=\langle G|,~~
\lim_{\eta \to 0}|G_\eta \rangle=|G \rangle,~~
\lim_{\eta \to 0}\langle F_\eta|=\langle F|,~~
\lim_{\eta \to 0}|F_\eta \rangle=|F \rangle.
\end{eqnarray}
First we compute the normal-ordering
of the vertex operators.\\
Let us denote by $A$ the index set
\begin{eqnarray}
A=\{a|\epsilon_a=-, 1\leq a \leq N\}
\end{eqnarray}
We have the following expressions.
\begin{eqnarray}
&&\frac{\langle G_\eta|
\Phi_{\epsilon_N}(\beta_N)\cdots \Phi_{\epsilon_1}(\beta_1)
|G_\eta\rangle
}{\langle G_\eta |G_\eta \rangle}
\nonumber\\
&=&
\prod_{1\leq b_2<b_1 \leq N}
\frac{\Gamma_2\left(i(\beta_{b_2}-\beta_{b_1})+2\pi
|\omega_1 \omega_2\right)
\Gamma_2\left(i(\beta_{b_2}-\beta_{b_1})+\pi(\xi+1)
|\omega_1 \omega_2\right)
}{
\Gamma_2\left(i(\beta_{b_2}-\beta_{b_1})+\pi
|\omega_1 \omega_2\right)
\Gamma_2\left(i(\beta_{b_2}-\beta_{b_1})+\pi(\xi+2)
|\omega_1 \omega_2\right)
}\nonumber\\
&\times&
\prod_{a \in A}\int_{-\infty}^\infty d\alpha_a
\prod_{a \in A}\Gamma\left(
\frac{i(\alpha_a-\beta_a)}{\pi(\xi+1)}+\frac{1}{2(\xi+1)}
\right)\Gamma\left(
\frac{i(\beta_a-\alpha_a)}{\pi(\xi+1)}+\frac{1}{2(\xi+1)}
\right)\nonumber\\
&\times&
\prod_{a_2<a_1 \atop{a_1,a_2 \in A}}
\left\{
(\alpha_{a_2}-\alpha_{a_1})
\frac{\displaystyle
\Gamma\left(\frac{i(\alpha_{a_2}-\alpha_{a_1})}{\pi(\xi+1)}
+1-\frac{1}{\xi+1}\right)}{
\displaystyle
\Gamma\left(\frac{i(\alpha_{a_2}-\alpha_{a_1})}{\pi(\xi+1)}
+\frac{1}{\xi+1}\right)
}\right\}\nonumber\\
&\times&
\prod_{a>b \atop{a \in A, 1\leq b \leq N}}
\frac{\displaystyle
\Gamma\left(\frac{i(\beta_b-\alpha_a)}{\pi(\xi+1)}+
\frac{1}{2(\xi+1)}\right)}{
\displaystyle
\Gamma\left(\frac{i(\beta_b-\alpha_a)}{\pi(\xi+1)}+
1-\frac{1}{2(\xi+1)}\right)
}
\prod_{b>a \atop{a \in A, 1\leq b \leq N}}
\frac{\displaystyle
\Gamma\left(\frac{i(\alpha_a-\beta_b)}{\pi(\xi+1)}+
\frac{1}{2(\xi+1)}\right)}{
\displaystyle
\Gamma\left(\frac{i(\alpha_a-\beta_b)}{\pi(\xi+1)}+
1-\frac{1}{2(\xi+1)}\right)
}\nonumber
\\
&\times&
I_\eta(\{\beta_b\}|\{\alpha_a\}).
\end{eqnarray}
Let us denote by $A^*$ the undex set
\begin{eqnarray}
A^*=\{a|\epsilon_a^*=+, 1\leq a \leq N\}
\end{eqnarray}
We have the following expressions.
\begin{eqnarray}
&&\frac{\langle F_\eta|
\Phi_{\epsilon_1*}^*(\beta_1^*-\pi i)
\cdots \Phi_{\epsilon_N*}(\beta_N^*-\pi i)
|F_\eta\rangle
}{\langle F_\eta |F_\eta \rangle}
\nonumber\\
&=&
\prod_{1\leq b_1<b_2 \leq N}
\frac{\Gamma_2\left(i(\beta_{b_2}-\beta_{b_1})+2\pi
|\omega_1 \omega_2\right)
\Gamma_2\left(i(\beta_{b_2}-\beta_{b_1})+\pi(\xi+1)
|\omega_1 \omega_2\right)
}{
\Gamma_2\left(i(\beta_{b_2}-\beta_{b_1})+\pi
|\omega_1 \omega_2\right)
\Gamma_2\left(i(\beta_{b_2}-\beta_{b_1})+\pi(\xi+2)
|\omega_1 \omega_2\right)
}\nonumber\\
&\times&
\prod_{a \in A^*}\int_{-\infty}^\infty d\alpha_a
\prod_{a \in A^*}\Gamma\left(
\frac{i(\alpha_a-\beta_a)}{\pi(\xi+1)}+\frac{1}{2(\xi+1)}
\right)\Gamma\left(
\frac{i(\beta_a-\alpha_a)}{\pi(\xi+1)}+\frac{1}{2(\xi+1)}
\right)\nonumber
\\
&\times&
\prod_{a_1<a_2 \atop{a_1,a_2 \in A^*}}
\left\{
(\alpha_{a_2}-\alpha_{a_1})
\frac{\displaystyle
\Gamma\left(\frac{i(\alpha_{a_2}-\alpha_{a_1})}{\pi(\xi+1)}
+1-\frac{1}{\xi+1}\right)}{
\displaystyle
\Gamma\left(\frac{i(\alpha_{a_2}-\alpha_{a_1})}{\pi(\xi+1)}
+\frac{1}{\xi+1}\right)
}\right\}\nonumber\\
&\times&
\prod_{a<b \atop{a \in A^*, 1\leq b \leq N}}
\frac{\displaystyle
\Gamma\left(\frac{i(\beta_b-\alpha_a)}{\pi(\xi+1)}+
\frac{1}{2(\xi+1)}\right)}{
\displaystyle
\Gamma\left(\frac{i(\beta_b-\alpha_a)}{\pi(\xi+1)}+
1-\frac{1}{2(\xi+1)}\right)
}
\prod_{b<a \atop{a \in A^*, 1\leq b \leq N}}
\frac{\displaystyle
\Gamma\left(\frac{i(\alpha_a-\beta_b)}{\pi(\xi+1)}+
\frac{1}{2(\xi+1)}\right)}{
\displaystyle
\Gamma\left(\frac{i(\alpha_a-\beta_b)}{\pi(\xi+1)}+
1-\frac{1}{2(\xi+1)}\right)
}\nonumber
\\
&\times&
I_\eta^*(\{\beta_b^*-\pi i\}|\{\alpha_a\}).
\end{eqnarray}
Here we have set
\begin{eqnarray}
I_\eta(\{\beta_b\}|\{\alpha_a\})=
\frac{\langle G_\eta|\exp\left(\int_0^\infty
X_A(t)b(-t)dt\right)
\exp\left(\int_0^\infty
Y_A(t)b(t)dt\right)|G_\eta\rangle
}{
\langle G_\eta |G_\eta\rangle
},\\
I^*_\eta(\{\beta_b^*-\pi i\}|\{\alpha_a^\})
=
\frac{\langle F_\eta|\exp\left(\int_0^\infty
X_{A^*}(t)b(-t)dt\right)
\exp\left(\int_0^\infty
Y_{A^*}(t)b(t)dt\right)|F_\eta\rangle
}{
\langle F_\eta |F_\eta\rangle
}.
\end{eqnarray}
with
\begin{eqnarray}
X_A(t)&=&\sum_{b=1}^N
\frac{e^{-i\beta_b t}}{{\rm sh}(\pi t)}-
\sum_{a \in A}\frac{e^{-i\alpha_a t}}{{\rm sh}
\left(\frac{\pi t}{2}\right)},\\
Y_A(t)&=&-
\sum_{b=1}^N
\frac{e^{i\beta_b t}}{{\rm sh}(\pi t)}+
\sum_{a \in A}\frac{e^{i\alpha_a t}}{{\rm sh}
\left(\frac{\pi t}{2}\right)}.
\end{eqnarray}
We evaluate the quantities 
$I_\eta(\{\beta_b\}|\{\alpha_a\}),
I_\eta^*(\{\beta_b^*-\pi i\}|\{\alpha_a\})
$.
Using the completeness relation of the coherent state
\cite{JKKKM, K}, and performing the integral calculations, 
we have
\begin{eqnarray}
&&I_\eta(\{\beta_b\}|\{\alpha_a\})
\nonumber\\
&=&
\exp\left(\int_0^\infty
\frac{1}{1-e^{-2\eta t}}
\frac{{\rm sh}\left(\frac{\pi t}{2}\right)
{\rm sh}\left(\pi t\right)
{\rm sh}\left(\frac{\pi \xi t}{2}\right)
}{t
{\rm sh}\left(\frac{\pi}{2}(\xi+1)t\right)
}
\right.\nonumber\\
&\times&
\left(-\frac{1}{2}e^{-\eta t} X_A(t)^2
+e^{-2\eta t}X_A(t)Y_A(t)-\frac{1}{2}e^{-\eta t}Y_A(t)^2
\right)\\
&+&\left.
\int_0^\infty
\frac{1}{1-e^{-2\eta t}}\{
(G_1(t|\mu)-e^{-\eta t}G_1^*(t|\nu))X_A(t)+
(G_1^*(t|\nu)-e^{-\eta t}G_1(t|\mu))Y_A(t)
\}dt
\right).\nonumber
\end{eqnarray}
and
\begin{eqnarray}
&&I_\eta^*(\{\beta_b^*-\pi i\}|\{\alpha_a\})
\nonumber\\
&=&
\exp\left(\int_0^\infty
\frac{1}{1-e^{-2\eta t}}
\frac{{\rm sh}\left(\frac{\pi t}{2}\right)
{\rm sh}\left(\pi t\right)
{\rm sh}\left(\frac{\pi \xi t}{2}\right)
}{t
{\rm sh}\left(\frac{\pi}{2}(\xi+1)t\right)
}
\right.\nonumber\\
&\times&
\left(-\frac{1}{2}e^{-\eta t-2\pi t} X_{A^*}(t)^2
+e^{-2\eta t}X_{A^*}(t)Y_{A^*}(t)
-\frac{1}{2}e^{-\eta t+2\pi t}Y_{A^*}(t)^2
\right)\\
&+&\left.
\int_0^\infty
\frac{1}{1-e^{-2\eta t}}\{
(F_1(t|\mu)-e^{-\eta t-2\pi t}F_1^*(t|\nu))X_{A^*}(t)+
(F_1^*(t|\nu)-e^{-\eta t+2\pi t}F_1(t|\mu))Y_{A^*}(t)
\}dt
\right).\nonumber
\end{eqnarray}

In what follows we use
the abberiviations :
\begin{eqnarray}
\omega_1=2\pi, 
\omega_2=\pi(\xi+1), \omega_3=2 \eta,~~
\mu_+=\mu, \mu_-=\nu,
\end{eqnarray}
The vacuum expectation 
value is evaluated as following.
\begin{eqnarray}
I_\eta(\{\beta_b\}|\{\alpha_a\})
=
I_\eta^\beta(\{\beta_b\})
I_\eta^{\beta \alpha}(\{\beta_b\}|\{\alpha_a\})
I_\eta^\alpha(\{\alpha_a\}).
\end{eqnarray}

Here we set
\begin{eqnarray}
&&I_\eta^\beta(\{\beta_b\})\nonumber\\
&=&\prod_{b=1}^{N}
\prod_{\epsilon=\pm}
\sqrt{\frac{S_3(2i\epsilon \beta_b
+\pi+\eta|\omega_1 \omega_2 \omega_3)}{
S_3(2i\epsilon \beta_b
+2\pi+\eta|\omega_1 \omega_2 \omega_3)
}}
\prod_{b_1<b_2}\prod_{\epsilon=\pm}
\frac{
\Gamma_2(i\epsilon(\beta_{b_1}-\beta_{b_2})+\pi
|\omega_1 \omega_2)
}{
\Gamma_2(i\epsilon(\beta_{b_1}-\beta_{b_2})+2\pi
|\omega_1 \omega_2)}
\nonumber\\
&\times&
\prod_{b_1<b_2}\prod_{\epsilon=\pm}
\frac{S_3(i\epsilon(\beta_{b_1}+\beta_{b_2})+\pi+\eta
|\omega_1 \omega_2 \omega_3)
S_3(i\epsilon(\beta_{b_1}-\beta_{b_2})+\pi
|\omega_1 \omega_2 \omega_3)
}{
S_3(i\epsilon(\beta_{b_1}+\beta_{b_2})+2\pi+\eta
|\omega_1 \omega_2 \omega_3)
S_3(i\epsilon(\beta_{b_1}-\beta_{b_2})+2\pi
|\omega_1 \omega_2 \omega_3)
}\nonumber
\\
&\times&
\prod_{b=1}^N\prod_{\epsilon=\pm}
\frac{
\Gamma_3(i\epsilon\beta_b+i\mu_\epsilon+\pi|
\omega_1\omega_2\omega_3)
\Gamma_3(i\epsilon\beta_b-i\mu_\epsilon
+\pi\xi+2\pi|\omega_1\omega_2\omega_3)
}{
\Gamma_3(i\epsilon\beta_b-i\mu_\epsilon+\pi\xi+\pi
|\omega_1\omega_2\omega_3)
\Gamma_3(i\epsilon\beta_b+
i\mu_\epsilon+2\pi|
\omega_1\omega_2\omega_3)
}\nonumber\\
&\times&
\prod_{b=1}^N\prod_{\epsilon=\pm}
\frac{
\Gamma_3(-i\epsilon\beta_b+\eta
+i\mu_\epsilon+\pi|\omega_1\omega_2\omega_3)
\Gamma_3(-i\epsilon\beta_b+\eta
-i\mu_\epsilon
+\pi\xi+2\pi|\omega_1\omega_2\omega_3)
}{
\Gamma_3(-i\epsilon\beta_b+\eta
-i\mu_\epsilon+\pi\xi+\pi|
\omega_1\omega_2\omega_3)
\Gamma_3(-i\epsilon\beta_b+\eta+
i\mu_\epsilon+2\pi|
\omega_1\omega_2\omega_3)
}\nonumber\\
&\times&
\prod_{b=1}^N\prod_{\epsilon=\pm}
\sqrt{\frac{\Gamma_3(2i\epsilon \beta_b+\pi|2\omega_1,\omega_2,
2\omega_3)
\Gamma_3(2i\epsilon \beta_b+4\pi|2\omega_1,\omega_2,
2\omega_3)
}{
\Gamma_3(2i\epsilon \beta_b+2\pi|2\omega_1,\omega_2,
2\omega_3)
\Gamma_3(2i\epsilon \beta_b+3\pi|2\omega_1,\omega_2,
2\omega_3)
}}\nonumber
\\
&\times&
\prod_{b=1}^N\prod_{\epsilon=\pm}
\sqrt{\frac{\Gamma_3(2i\epsilon \beta_b+\pi\xi+\pi|
2\omega_1,\omega_2,
2\omega_3)
\Gamma_3(2i\epsilon \beta_b+\pi\xi+4\pi|2\omega_1,\omega_2,
2\omega_3)
}{
\Gamma_3(2i\epsilon \beta_b+\pi\xi+2\pi|2\omega_1,\omega_2,
2\omega_3)
\Gamma_3(2i\epsilon \beta_b+\pi\xi+3\pi|2\omega_1,\omega_2,
2\omega_3)
}}\nonumber
\\
&\times&
\prod_{b=1}^N\prod_{\epsilon=\pm}
\sqrt{\frac{\Gamma_3(2i\epsilon \beta_b+2\eta+\pi|2\omega_1,\omega_2,
2\omega_3)
\Gamma_3(2i\epsilon \beta_b+2\eta+4\pi|2\omega_1,\omega_2,
2\omega_3)
}{
\Gamma_3(2i\epsilon \beta_b+2\eta+2\pi|2\omega_1,\omega_2,
2\omega_3)
\Gamma_3(2i\epsilon \beta_b+2\eta+3\pi|2\omega_1,\omega_2,
2\omega_3)
}}
\\
&\times&
\prod_{b=1}^N\prod_{\epsilon=\pm}
\sqrt{\frac{\Gamma_3(2i\epsilon \beta_b+2\eta+\pi\xi+\pi|
2\omega_1,\omega_2,
2\omega_3)
\Gamma_3(2i\epsilon \beta_b+2\eta+\pi\xi+4\pi|2\omega_1,\omega_2,
2\omega_3)
}{
\Gamma_3(2i\epsilon \beta_b+2\eta+\pi\xi+2\pi|2\omega_1,\omega_2,
2\omega_3)
\Gamma_3(2i\epsilon \beta_b+2\eta+
\pi\xi+3\pi|2\omega_1,\omega_2,
2\omega_3)
}}\nonumber
\end{eqnarray}

\begin{eqnarray}
&&I_\eta^\alpha(\{\alpha_a\})\nonumber\\
&=&
\prod_{a \in A}\prod_{\epsilon=\pm}
\sqrt{\frac{S_2(2i\epsilon \alpha_a+\eta|\omega_2 \omega_3)}{
S_2(2i\epsilon \alpha_a+\pi \xi+\eta|\omega_2 \omega_3)
}}
\prod_{a_1<a_2}
\prod_{\epsilon=\pm}
\frac{\Gamma\left(
\frac{i(\alpha_{a_1}-\alpha_{a_2})}{\pi(\xi+1)}+\frac{1}{\xi+1}
\right)}{
\Gamma\left(
\frac{i(\alpha_{a_1}-\alpha_{a_2})}{\pi(\xi+1)}+1
\right)}
\nonumber\\
&\times&\prod_{a_1<a_2}
\prod_{\epsilon=\pm}
\frac{S_2(i\epsilon(\alpha_{a_1}+\alpha_{a_2})+\eta|
\omega_2 \omega_3)
S_2(i\epsilon(\alpha_{a_1}-\alpha_{a_2})+\pi|
\omega_2 \omega_3)
}{
S_2(i\epsilon(\alpha_{a_1}+\alpha_{a_2})+\pi \xi+\eta|
\omega_2 \omega_3)
S_2(i\epsilon(\alpha_{a_1}-\alpha_{a_2})+\pi(\xi+1)|
\omega_2 \omega_3)
}\nonumber\\
&\times&
\prod_{a \in A}\prod_{\epsilon=\pm}
\frac{\Gamma_2(i\epsilon \alpha_a+\pi \xi
+\frac{\pi}{2}-i\mu_\epsilon
|\omega_2 \omega_3)
\Gamma_2(-i\epsilon \alpha_a+\eta+\pi \xi
+\frac{\pi}{2}-i\mu_\epsilon
|\omega_2 \omega_3)
}{
\Gamma_2(i\epsilon \alpha_a+\frac{\pi}{2}
+i\mu_{\epsilon}|\omega_2 \omega_3)
\Gamma_2(
-i\epsilon \alpha_a+\eta+\frac{\pi}{2}
+i\mu_{\epsilon}
|\omega_2 \omega_3)
}\nonumber\\
&\times&
\prod_{a \in A}\prod_{\epsilon=\pm}
\sqrt{\frac{
\Gamma_2(
2i\epsilon \alpha_a+\pi
|\omega_2,2\omega_3)
\Gamma_2(
2i\epsilon \alpha_a+\pi(\xi+1)
|\omega_2,2\omega_3)
}{
\Gamma_2(
2i\epsilon \alpha_a
|\omega_2,2\omega_3)
\Gamma_2(
2i\epsilon \alpha_a+\pi \xi
|\omega_2,2\omega_3)
}}
\\
&\times&
\prod_{a \in A}
\prod_{\epsilon=\pm}
\sqrt{\frac{
\Gamma_2(
2i\epsilon \alpha_a+2\eta+\pi
|\omega_2,2\omega_3)
\Gamma_2(
2i\epsilon \alpha_a+2\eta+\pi(\xi+1)
|\omega_2,2\omega_3)
}{
\Gamma_2(
2i\epsilon \alpha_a+2\eta
|\omega_2,2\omega_3)
\Gamma_2(
2i\epsilon \alpha_a+2\eta+\pi \xi
|\omega_2,2\omega_3)
}}.\nonumber
\end{eqnarray}

\begin{eqnarray}
&&I_\eta^{\beta \alpha}(\{\beta_b\}|\{\alpha_a\})
\nonumber\\
&=&
\prod_{a \in A}\prod_{b=1}^N
\frac{S_2(i(\alpha_a+\beta_b)+\eta+\pi \xi +\frac{\pi}{2}
|\omega_2 \omega_3)}{
S_2(i(\alpha_a+\beta_b)+\eta+\frac{\pi}{2}
|\omega_2 \omega_3)
}\\
&\times&
\prod_{a \in A}\prod_{b=1}^N
\prod_{\epsilon=\pm}
\left\{\Gamma\left(\frac{i\epsilon(\alpha_a-\beta_b)}
{\pi(\xi+1)}+\frac{1}{2(\xi+1)}\right)
S_2\left.\left(
i\epsilon(\alpha_a-\beta_b)+\frac{\pi}{2}\right|\omega_2 \omega_3
\right)
\right\}^{-1}.\nonumber
\end{eqnarray}
The vacuum expectation value is evaluated as following.
\begin{eqnarray}
I_\eta^*(\{\beta_b^*-\pi i\}|\{\alpha_a\})
=
I_\eta^{*\beta}(\{\beta_b^*\})
I_\eta^{*\beta \alpha}(\{\beta_b^*\}|\{\alpha_a\})
I_\eta^{*\alpha}(\{\alpha_a\}).
\end{eqnarray}
Here we set

\begin{eqnarray}
&&I_\eta^{*\beta}(\{\beta_b^*\})
\nonumber\\
&=&\prod_{b=1}^{N}
\prod_{\epsilon=\pm}
\sqrt{\frac{S_3(2i\epsilon \beta_b^*
+\pi+2\pi\epsilon+\eta|\omega_1 \omega_2 \omega_3)}{
S_3(2i\epsilon \beta_b^*
+2\pi+2\pi\epsilon+\eta|\omega_1 \omega_2 \omega_3)
}}
\prod_{b_1<b_2}\prod_{\epsilon=\pm}
\frac{
\Gamma_2(i\epsilon(\beta_{b_1}^*-\beta_{b_2}^*)+\pi
|\omega_1 \omega_2)
}{
\Gamma_2(i\epsilon(\beta_{b_1}^*-\beta_{b_2}^*)+2\pi
|\omega_1 \omega_2)}
\nonumber\\
&\times&
\prod_{b_1<b_2}\prod_{\epsilon=\pm}
\frac{S_3(i\epsilon(\beta_{b_1}^*+\beta_{b_2}^*)+\pi+
2\pi\epsilon+\eta
|\omega_1 \omega_2 \omega_3)
S_3(i\epsilon(\beta_{b_1}^*-\beta_{b_2}^*)+\pi
|\omega_1 \omega_2 \omega_3)
}{
S_3(i\epsilon(\beta_{b_1}^*+\beta_{b_2}^*)+2\pi+
2\pi\epsilon+\eta
|\omega_1 \omega_2 \omega_3)
S_3(i\epsilon(\beta_{b_1}^*-\beta_{b_2}^*)+2\pi
|\omega_1 \omega_2 \omega_3)
}\nonumber
\\
&\times&\prod_{b=1}^N
\prod_{\epsilon=\pm}
\frac{\Gamma_3(
i\epsilon\beta_b^*-i\mu_\epsilon+\pi\xi+2\pi+\pi\epsilon
|\omega_1\omega_2\omega_3)
\Gamma_3(
i\epsilon\beta_b^*+i\mu_\epsilon+\pi\epsilon+\pi
|\omega_1\omega_2\omega_3)
}{
\Gamma_3(
i\epsilon\beta_b^*+i\mu_\epsilon+\pi\epsilon
|\omega_1\omega_2\omega_3)
\Gamma_3(
i\epsilon\beta_b^*-i\mu_\epsilon+\pi\xi+3\pi+\pi\epsilon
|\omega_1\omega_2\omega_3)
}
\nonumber
\\
&\times&\prod_{b=1}^N
\prod_{\epsilon=\pm}
\frac{
\Gamma_3(
-i\epsilon\beta_b^*-i\mu_\epsilon+\eta+\pi\xi+2\pi-\pi\epsilon
|\omega_1\omega_2\omega_3)
\Gamma_3(
-i\epsilon\beta_b^*+\eta+i\mu_\epsilon+\pi-\pi\epsilon
|\omega_1\omega_2\omega_3)
}{
\Gamma_3(
-i\epsilon \beta_b^*+i\mu_\epsilon
+\eta-\pi\epsilon|\omega_1\omega_2\omega_3)
\Gamma_3(-i\epsilon\beta_b^*
-i\mu_\epsilon+\eta+\pi\xi+\pi-\pi\epsilon
|\omega_1\omega_2\omega_3)
}\nonumber\\
&\times&
\prod_{b=1}^N\prod_{\epsilon=\pm}
\sqrt{\frac{\Gamma_3(2i\epsilon \beta_b^*+
2\pi\epsilon+\pi|2\omega_1,\omega_2,
2\omega_3)
\Gamma_3(2i\epsilon \beta_b^*+
2\pi\epsilon+4\pi|2\omega_1,\omega_2,
2\omega_3)
}{
\Gamma_3(2i\epsilon \beta_b^*+
2\pi\epsilon+2\pi|2\omega_1,\omega_2,
2\omega_3)
\Gamma_3(2i\epsilon \beta_b^*+
2\pi\epsilon+3\pi|2\omega_1,\omega_2,
2\omega_3)
}}\nonumber
\\
&\times&
\prod_{b=1}^N\prod_{\epsilon=\pm}
\sqrt{\frac{\Gamma_3(2i\epsilon \beta_b^*+2\pi\epsilon+\pi\xi+\pi|
2\omega_1,\omega_2,
2\omega_3)
\Gamma_3(2i\epsilon \beta_b^*+2\pi\epsilon+
\pi\xi+4\pi|2\omega_1,\omega_2,
2\omega_3)
}{
\Gamma_3(2i\epsilon \beta_b^*+2\pi\epsilon+
\pi\xi+2\pi|2\omega_1,\omega_2,
2\omega_3)
\Gamma_3(2i\epsilon \beta_b^*+2\pi\epsilon+
\pi\xi+3\pi|2\omega_1,\omega_2,
2\omega_3)
}}\nonumber
\\
&\times&
\prod_{b=1}^N\prod_{\epsilon=\pm}
\sqrt{\frac{\Gamma_3(2i\epsilon \beta_b^*+
2\pi\epsilon+2\eta+\pi|2\omega_1,\omega_2,
2\omega_3)
\Gamma_3(2i\epsilon \beta_b^*+2\pi\epsilon+
2\eta+4\pi|2\omega_1,\omega_2,
2\omega_3)
}{
\Gamma_3(2i\epsilon \beta_b^*+2\pi\epsilon+
2\eta+2\pi|2\omega_1,\omega_2,
2\omega_3)
\Gamma_3(2i\epsilon \beta_b^*+2\pi\epsilon+
2\eta+3\pi|2\omega_1,\omega_2,
2\omega_3)
}}\nonumber
\\
&\times&
\prod_{b=1}^N\prod_{\epsilon=\pm}
\sqrt{\frac{\Gamma_3(2i\epsilon \beta_b^*+2\pi\epsilon+
2\eta+\pi\xi+\pi|
2\omega_1,\omega_2,
2\omega_3)
}{
\Gamma_3(2i\epsilon \beta_b^*+2\pi\epsilon+
2\eta+\pi\xi+2\pi|2\omega_1,\omega_2,
2\omega_3)
}}\nonumber\\
&\times&
\prod_{b=1}^N\prod_{\epsilon=\pm}
\sqrt{\frac{
\Gamma_3(2i\epsilon \beta_b^*+2\pi\epsilon+
2\eta+\pi\xi+4\pi|2\omega_1,\omega_2,
2\omega_3)
}{
\Gamma_3(2i\epsilon \beta_b^*+2\pi\epsilon+2\eta+
\pi\xi+3\pi|2\omega_1,\omega_2,
2\omega_3)
}}.
\end{eqnarray}

\begin{eqnarray}
&&I_\eta^{*\beta \alpha}(\{\beta_b^*\}|\{\alpha_a\})
\nonumber\\
&=&
\prod_{a \in A^*}\prod_{b=1}^N
\frac{S_2(i(\alpha_a+\beta_b^*)+\eta+\pi \xi -\frac{3\pi}{2}
|\omega_2 \omega_3)}{
S_2(i(\alpha_a+\beta_b^*)+\eta-\frac{3\pi}{2}
|\omega_2 \omega_3)
}\\
&\times&
\prod_{a \in A^*}\prod_{b=1}^N
\prod_{\epsilon=\pm}
\left\{\Gamma\left(\frac{i\epsilon(\alpha_a-\beta_b^*)}
{\pi(\xi+1)}+\frac{1}{2(\xi+1)}\right)
S_2\left.\left(
i\epsilon(\alpha_a-\beta_b^*)+
\frac{\pi}{2}\right|\omega_2 \omega_3
\right)
\right\}^{-1}.\nonumber
\end{eqnarray}

\begin{eqnarray}
&&I_\eta^{*\alpha}
(\{\alpha_a\})\nonumber\\
&=&
\prod_{a \in A^*}\prod_{\epsilon=\pm}
\sqrt{\frac{S_2(2i\epsilon \alpha_a+2\pi\epsilon+
\eta|\omega_2 \omega_3)}{
S_2(2i\epsilon \alpha_a+\pi \xi+2\pi\epsilon+
\eta|\omega_2 \omega_3)
}}
\prod_{a_1<a_2}
\prod_{\epsilon=\pm}
\frac{\Gamma\left(
\frac{i(\alpha_{a_1}-\alpha_{a_2})}{\pi(\xi+1)}+\frac{1}{\xi+1}
\right)}{
\Gamma\left(
\frac{i(\alpha_{a_1}-\alpha_{a_2})}{\pi(\xi+1)}+1
\right)}
\nonumber\\
&\times&\prod_{a_1<a_2}
\prod_{\epsilon=\pm}
\frac{S_2(i\epsilon(\alpha_{a_1}+\alpha_{a_2})+2\pi\epsilon+
\eta|
\omega_2 \omega_3)
S_2(i\epsilon(\alpha_{a_1}-\alpha_{a_2})+\pi|
\omega_2 \omega_3)
}{
S_2(i\epsilon(\alpha_{a_1}+\alpha_{a_2})+\pi \xi+2\pi\epsilon+
\eta|
\omega_2 \omega_3)
S_2(i\epsilon(\alpha_{a_1}-\alpha_{a_2})+\pi(\xi+1)|
\omega_2 \omega_3)
}\nonumber\\
&\times&
\prod_{a \in A}\prod_{\epsilon=\pm}
\frac{\Gamma_2(i\epsilon \alpha_a+
i\mu_\epsilon-\frac{\pi}{2}+\pi\epsilon
|\omega_2 \omega_3)
\Gamma_2(-i\epsilon \alpha_a+
\eta-\frac{\pi}{2}-\pi\epsilon+i\mu_\epsilon
|\omega_2 \omega_3)
}{
\Gamma_2(i\epsilon \alpha_a-i\mu_\epsilon
+\pi\xi+\frac{\pi}{2}+\pi\epsilon|\omega_2 \omega_3)
\Gamma_2(
-i\epsilon \alpha_a+\eta+\pi\xi+\frac{\pi}{2}-\pi\epsilon-
i\mu_\epsilon
|\omega_2 \omega_3)
}\nonumber\\
&\times&
\prod_{a \in A}\prod_{\epsilon=\pm}
\sqrt{\frac{
\Gamma_2(
2i\epsilon \alpha_a+2\pi\epsilon+\pi
|\omega_2,2\omega_3)
\Gamma_2(
2i\epsilon \alpha_a+2\pi\epsilon+\pi(\xi+1)
|\omega_2,2\omega_3)
}{
\Gamma_2(
2i\epsilon \alpha_a+2\pi\epsilon
|\omega_2,2\omega_3)
\Gamma_2(
2i\epsilon \alpha_a+2\pi\epsilon+\pi \xi
|\omega_2,2\omega_3)
}}
\\
&\times&
\prod_{a \in A}
\prod_{\epsilon=\pm}
\sqrt{\frac{
\Gamma_2(
2i\epsilon \alpha_a+2\pi\epsilon+2\eta+\pi
|\omega_2,2\omega_3)
\Gamma_2(
2i\epsilon \alpha_a+2\pi\epsilon+2\eta+\pi(\xi+1)
|\omega_2,2\omega_3)
}{
\Gamma_2(
2i\epsilon \alpha_a+2\pi\epsilon+2\eta
|\omega_2,2\omega_3)
\Gamma_2(
2i\epsilon \alpha_a+2\pi\epsilon+2\eta+\pi \xi
|\omega_2,2\omega_3)
}}.\nonumber
\end{eqnarray}

~\\
The magnetization on a site $m$ is given by
\begin{eqnarray}
\langle \sigma_m^z \rangle=
\frac{\displaystyle
\sum_{\epsilon_1, \cdots ,\epsilon_N=\pm}
\epsilon_m
P_{\epsilon_1,\cdots,\epsilon_N;
\epsilon_N,\cdots,\epsilon_1;0}(\{0\}|\{0\})
}{\displaystyle
\sum_{\epsilon_1, \cdots ,\epsilon_N=\pm}
P_{\epsilon_1,\cdots,\epsilon_N;
\epsilon_N,\cdots,\epsilon_1;0}(\{0\}|\{0\})
}.
\end{eqnarray}

~\\
{\it Note.~
In paper \cite{FW} the authors considered 
the following vacuum expectation value
for finite XXZ chain with double
boundaries at massive regime,}
\begin{eqnarray}
\frac{
\langle vac |e^F \Phi_{\epsilon_1^*}(\zeta_1^*)
\cdots \Phi_{\epsilon_N^*}(\zeta_N^*) e^{F^*}e^{G}
\Phi_{\epsilon_N}(\zeta_N)\cdots \Phi_{\epsilon_1}(\zeta_1)
e^{G^*}|vac\rangle}
{\langle vac |e^{F}e^{F^*}e^{G}e^{G^*}|vac
\rangle},\nonumber
\end{eqnarray}
{\it
which is free from a difficulty of divergence.
However it's physical meaning is not clear.
}

~\\
{\bf Acknowledgements}~~This work was partly supported by
Grant-in-Aid for Encouragements for Young Scientists (A)
from Japan Society for the Promotion of Science (11740099).

\begin{appendix}
\section{Vertex Operators}
Here we summarize the bosonizations
of the vertex operators \cite{JKM}.\\
Let us set free bosons $b(t) (t \in {\mathbb{R}})$
which satisfy
\begin{eqnarray}
[b(t),b(t')]=\frac{\displaystyle
{\rm sh}\left(\frac{\pi t}{2}\right) 
{\rm sh}(\pi t) 
{\rm sh}\frac{\pi t \xi}{2} 
}{
\displaystyle
t {\rm sh}\frac{\pi t (\xi+1)}{2} }\delta(t+t').
\end{eqnarray}
Let us set $a(t)$ by
\begin{eqnarray}
b(t){\rm sh}\frac{\pi t (\xi+1)}{2}=
a(t){\rm sh}\frac{\pi t \xi}{2}.
\end{eqnarray}
The bosonization of the type-I vertex operators
is given by
\begin{eqnarray}
\Phi_+(\beta)&=&U(\beta),\\
\Phi_-(\beta)&=&\int_{C_I}d\alpha
:U(\beta)\bar{U}(\alpha):\nonumber\\
&\times&
\Gamma\left(\frac{i(\alpha-\beta)}{\pi(\xi+1)}+
\frac{1}{2(\xi+1)}\right)
\Gamma\left(-\frac{i(\alpha-\beta)}{\pi(\xi+1)}+
\frac{1}{2(\xi+1)}\right),
\end{eqnarray}
where we have set
\begin{eqnarray}
U(\alpha)=:\exp\left(-
\int_{-\infty}^\infty \frac{b(t)}{{\rm sh} \pi t}
e^{i\alpha t}dt\right):,~~
\bar{U}(\alpha)=:\exp\left(
\int_{-\infty}^\infty \frac{b(t)}{{\rm sh} \frac{\pi}{2} t}
e^{i\alpha t}dt\right):.
\end{eqnarray}
The bosonization of the type-II vertex operators is
given by
\begin{eqnarray}
\Psi_+(\beta)&=&V(\beta),\\
\Psi_-(\beta)&=&\int_{C_{II}}d\alpha
:V(\beta)\bar{V}(\alpha):\nonumber\\
&\times&
\Gamma\left(\frac{i(\alpha-\beta)}{\pi \xi}-
\frac{1}{2 \xi}\right)
\Gamma\left(-\frac{i(\alpha-\beta)}{\pi \xi}+
-\frac{1}{2\xi}\right),
\end{eqnarray}
where we have set
\begin{eqnarray}
V(\alpha)=:exp\left(
\int_{-\infty}^\infty \frac{a(t)}{{\rm sh} \pi t}
e^{i\alpha t}dt\right):,~~
\bar{V}(\alpha)=:exp\left(
-\int_{-\infty}^\infty \frac{a(t)}{{\rm sh} \frac{\pi}{2} t}
e^{i\alpha t}dt\right):.
\end{eqnarray}
Here the integration contours are chosen as follows.
The contour $C_I$ is $(-\infty, \infty)$.
The poles
\begin{eqnarray}
\alpha-\beta=\frac{\pi i}{2}+n\pi(\xi+1)i,~~(n \in {\mathbb{N}})
\end{eqnarray}
of $\Gamma\left(\frac{i(\alpha-\beta)}{\pi(\xi+1)}+
\frac{1}{2(\xi+1)}\right)$ are above $C_I$
and
the poles
\begin{eqnarray}
\alpha-\beta=-\frac{\pi i}{2}-n\pi(\xi+1)i,~~(n \in {\mathbb{N}})
\end{eqnarray}
of $\Gamma\left(-\frac{i(\alpha-\beta)}{\pi(\xi+1)}+
\frac{1}{2(\xi+1)}\right)$ are below $C_I$.
The contour $C_{II}$ is $(-\infty, \infty)$
except that
the poles
\begin{eqnarray}
\alpha-\beta=-\frac{\pi i}{2}+n\pi \xi i,~~(n \in {\mathbb{N}})
\end{eqnarray}
of $\Gamma\left(\frac{i(\alpha-\beta)}{\pi \xi}-
\frac{1}{2 \xi}\right)$ are above $C_{II}$
and
the poles
\begin{eqnarray}
\alpha-\beta=\frac{\pi i}{2}-n\pi \xi i,~~(n \in {\mathbb{N}})
\end{eqnarray}
of $\Gamma\left(-\frac{i(\alpha-\beta)}{\pi \xi}-
\frac{1}{2 \xi}\right)$ are below $C_{II}$.

\section{Multi Gamma functions}

Here we summarize the multiple gamma and the multiple sine
functions.\\
Let us set the functions
$\Gamma_1(x|\omega), \Gamma_2(x|\omega_1, \omega_2)
$ and
$\Gamma_3(x|\omega_1, \omega_2, \omega_3)
$
by
\begin{eqnarray}
{\rm log}\Gamma_1(x|\omega)+\gamma B_{11}(x|\omega)&=&
\int_C\frac{dt}{2\pi i t}e^{-xt}
\frac{{\rm log}(-t)}{1-e^{-\omega t}},\\
{\rm log}\Gamma_2(x|\omega_1, \omega_2)
-\frac{\gamma}{2} B_{22}(x|\omega_1, \omega_2)&=&
\int_C\frac{dt}{2\pi i t}e^{-xt}
\frac{{\rm log}(-t)}{(1-e^{-\omega_1 t})
(1-e^{-\omega_2 t})},\\
{\rm log}\Gamma_3(x|\omega_1, \omega_2, \omega_3)
+\frac{\gamma}{3!} B_{33}(x|\omega_1, \omega_2, \omega_3)&=&
\int_C\frac{dt}{2\pi i t}e^{-xt}
\frac{{\rm log}(-t)}{(1-e^{-\omega_1 t})
(1-e^{-\omega_2 t})
(1-e^{-\omega_3 t})},\nonumber
\\
\end{eqnarray}
where
the functions $B_{jj}(x)$ are the multiple Bernoulli polynomials
defined by
\begin{eqnarray}
\frac{t^r e^{xt}}{
\prod_{j=1}^r (e^{\omega_j t}-1)}=
\sum_{n=0}^\infty
\frac{t^n}{n!}B_{r,n}(x|\omega_1 \cdots \omega_r),
\end{eqnarray}
more explicitly
\begin{eqnarray}
B_{11}(x|\omega)&=&\frac{x}{\omega}-\frac{1}{2},\\
B_{22}(x|\omega)&=&\frac{x^2}{\omega_1 \omega_2}
-\left(\frac{1}{\omega_1}+\frac{1}{\omega_2}\right)x
+\frac{1}{2}+\frac{1}{6}\left(\frac{\omega_1}{\omega_2}
+\frac{\omega_2}{\omega_1}\right).
\end{eqnarray}
Here $\gamma$ is Euler's constant,
$\gamma=\lim_{n\to \infty}
(1+\frac{1}{2}+\frac{1}{3}+\cdots+\frac{1}{n}-{\rm log}n)$.\\
Here the contor of integral is given by

~\\
~\\

\unitlength 0.1in
\begin{picture}(34.10,11.35)(17.90,-19.35)
%
\special{pn 8}%
\special{pa 5200 800}%
\special{pa 2190 800}%
\special{fp}%
\special{sh 1}%
\special{pa 2190 800}%
\special{pa 2257 820}%
\special{pa 2243 800}%
\special{pa 2257 780}%
\special{pa 2190 800}%
\special{fp}%
\special{pa 2190 1600}%
\special{pa 5190 1600}%
\special{fp}%
\special{sh 1}%
\special{pa 5190 1600}%
\special{pa 5123 1580}%
\special{pa 5137 1600}%
\special{pa 5123 1620}%
\special{pa 5190 1600}%
\special{fp}%
%
\special{pn 8}%
\special{pa 5190 1200}%
\special{pa 2590 1210}%
\special{fp}%
\put(25.9000,-12.1000){\makebox(0,0)[lb]{$0$}}%
%
\special{pn 8}%
\special{ar 2190 1210 400 400  1.5707963 4.7123890}%
\put(33.9000,-20.2000){\makebox(0,0){{\bf Contour} $C$}}%
\end{picture}%

~\\
~\\

Let us set
\begin{eqnarray}
S_1(x|\omega)&=&\frac{1}{\Gamma_1(\omega-x|\omega)
\Gamma_1(x|\omega)},\\
S_2(x|\omega_1,\omega_2)&=&\frac{
\Gamma_2(\omega_1+\omega_2-x|\omega_1,\omega_2)}{
\Gamma_2(x|\omega_1,\omega_2)},\\
S_3(x|\omega_1,\omega_2,\omega_3)&=&\frac{1}{
\Gamma_3(\omega_1+\omega_2+\omega_3-x|\omega_1,\omega_2,\omega_3)
\Gamma_3(x|\omega_1,\omega_2,\omega_3)}
\end{eqnarray}
We have
\begin{eqnarray}
\Gamma_1(x|\omega)=e^{(\frac{x}{\omega}-\frac{1}{2}){\rm log}
\omega}\frac{\Gamma(x/\omega)}{\sqrt{2\pi}},~
S_1(x|\omega)=2{\rm sin}(\pi x/\omega),
\end{eqnarray}
\begin{eqnarray}
\frac{\Gamma_2(x+\omega_1|\omega_1,\omega_2)}{
\Gamma_2(x|\omega_1,\omega_2)}=\frac{1}{\Gamma_1(x|\omega_2)},~
\frac{S_2(x+\omega_1|\omega_1,\omega_2)}{
S_2(x|\omega_1,\omega_2)}=\frac{1}{S_1(x|\omega_2)},~
\frac{\Gamma_1(x+\omega|\omega)}{\Gamma_1(x|\omega)}=x.
\end{eqnarray}

\begin{eqnarray}
\frac{
\Gamma_3(x+\omega_1|\omega_1,\omega_2, \omega_3)
}{\Gamma_3(x|\omega_1,\omega_2, \omega_3)}
=\frac{1}{\Gamma_2(x|\omega_2, \omega_3)},~
\frac{S_3(x+\omega_1|\omega_1,\omega_2,\omega_3)}{
S_3(x|\omega_1,\omega_2,\omega_3)}=\frac{1}{S_2(x|\omega_2,
\omega_3)}.
\end{eqnarray}

\begin{eqnarray}
{\rm log}S_2(x|\omega_1 \omega_2)
=\int_C \frac{{\rm sh}(x-\frac{\omega_1+\omega_2}{2})t}
{2{\rm sh}\frac{\omega_1 t}{2}
{\rm sh}\frac{\omega_2 t}{2}
}{\rm log}(-t)\frac{dt}{2\pi i t},~(0<{\rm Re}x<
\omega_1+\omega_2).
\end{eqnarray}

\begin{eqnarray}
S_2(x|\omega_1 \omega_2)=
\frac{2\pi}{\sqrt{\omega_1 \omega_2}}x +O(x^2),~~(x \to 0).
\end{eqnarray}

\begin{eqnarray}
S_2(x|\omega_1 \omega_2)
S_2(-x|\omega_1 \omega_2)=-4
{\rm sin}\frac{\pi x}{\omega_1}
{\rm sin}\frac{\pi x}{\omega_2}.
\end{eqnarray}
\end{appendix}
\end{document}